\input phyzzx
\hsize=40pc
                    %San-Serif 10
                   %San-Serif 8
                 %Bold San-Serif 10
    %Bold San-Serif 12

%\def\Figure#1#2{\midinsert
%$$\BoxedEPSF{#1}$$
%\noindent {\sf #2}
%\endinsert}

%\input BoxedEPS
%%\SetTexturesEPSFSpecial
%\SetRokickiEPSFSpecial %%for unix dvips
%\HideDisplacementBoxes 

\catcode`\@=11 % This allows us to modify PLAIN macros.
\def\NEWrefmark#1{\step@ver{{\;#1}}}
\catcode`\@=12 % at signs are no longer letters

\def\square{\kern1pt\vbox{\hrule height 1.2pt\hbox{\vrule width 1.2pt\hskip
3pt
   \vbox{\vskip 6pt}\hskip 3pt\vrule width 0.6pt}\hrule height 0.6pt}\kern1pt}
\def\inbar{\,\vrule height 1.5ex width .4pt depth0pt}\def\IC{\relax
\hbox{$\inbar\kern-.3em{\rm C}$}}
\def\IH{\relax\hbox{$\inbar\kern-.3em{\rm H}$}}

\def\bra#1{\langle #1 |}
\def\ket#1{| #1 \rangle}

\def\A{{\cal A}}
\def\B{{\cal B}}

\def\H{{\cal H}}

\def\I{{\cal I}}

\def\K{{\cal K}}

\def\L{{\cal L}}
\def\M{{\cal M}}

\def\O{{\cal O}}

\def\R{{\cal R}}

\def\T{{\cal T}}

\def\V{{\cal V}}

\def\p{\partial}

\def\wt{\widetilde}
\def\wh{\widehat}

\def\B{{\cal B}}

\def\V{{\cal V}}
\def\O{{\cal O}}

\def\p{\partial}

\def\X{{\cal X}}
\def\Y{{\cal Y}}
%%%%%%%%%%%%%%%%%%%%%%%%%%%%%%%%%%%%%%%%%%%%%%%%%%%%%%%%%%%%%%%%%%%%%%%%%%%%%
\singlespace

\def\define#1#2\par{\def#1{\Ref#1{#2}\edef#1{\noexpand\refmark{#1}}}}
\def\con#1#2\noc{\let\?=\Ref\let\<=\refmark\let\Ref=\REFS
         \let\refmark=\undefined#1\let\Ref=\REFSCON#2
         \let\Ref=\?\let\refmark=\<\refsend}

\let\refmark=\NEWrefmark

\define\csft{B. Zwiebach, `Closed string field theory: quantum action and the
Batalin-Vilkovisky master equation',
Nucl. Phys. {\bf B390} (1993) 33, hep-th/9206084.}

\define\manifest{S. Rahman, `The path to manifest background independence',
MIT-CTP-2649, {\it to appear.}}

\define\action{S. Rahman, `Geometrising the closed string field theory
action',
MIT-CTP-2648, {\it to appear.}}

\define\cbi{A. Sen and B. Zwiebach, `Local background independence of
classical closed string field theory',
Nucl. Phys. {\bf B414} (1994) 649, hep-th/9307088.}

\define\qbi{A. Sen and B. Zwiebach, `Quantum background independence of closed
string field theory',
Nucl. Phys. {\bf B423} (1994) 580, hep-th/9311009.}

\define\nonconf{B. Zwiebach, `Building string field theory around
non-conformal backgrounds',
Nucl. Phys. {\bf B480} (1996) 541, hep-th/9606153.}

\define\moduli{B. Zwiebach, `New moduli spaces from string background
independence consistency conditions',
Nucl. Phys. {\bf B480} (1996) 507, hep-th/9605075.}

\define\basisone{S. Rahman, `Consistency of quantum background independence',
MIT-CTP-2626, April 1997, hep-th/9704141.}

\define\bistruct{A. Sen and B. Zwiebach, `Background independent algebraic
structures in closed string field theory',
Comm. Math. Phys. {\bf 177} (1996) 305, hep-th/9408053.}

%%%%%%%%%%%%%%%%%%%%%%%%%%%%%%%%%%%%%%%%
\singlespace
%%%%%%%%%%%%%%%%%%%%%%%%%%%%%%%%%%%%%%%%%%%%%%%%%%%%%%%%%%%%%%%%%%%%%%%
%$$Bismillahirrahmanirraheem$$
{}~ \hfill \vbox{\hbox{MIT-CTP-2645}
%\hbox{hep-th/9706128}
\hbox{
} }\break
\title{STRING VERTICES AND INNER DERIVATIONS}
\author{Sabbir A Rahman \foot{E-mail address: rahman@marie.mit.edu
\hfill\break Supported in part by D.O.E.
cooperative agreement DE-FC02-94ER40818.}}
\address{Center for Theoretical Physics,\break
Laboratory of Nuclear Science\break
and Department of Physics\break
Massachusetts Institute of Technology\break
Cambridge, Massachusetts 02139, U.S.A.}

\abstract
{We show that it is algebraically consistent to express the string field
theory operators $\p$, $\K$ and $\I$ as inner derivations of the B-V algebra
of string vertices. In this approach, the recursion relations for the string
vertices are found to take the form of a `geometrical' quantum master
equation,
$\half \{ \B , \B \} + \Delta \B = 0$. We also show that the B-V delta
operator cannot be an inner derivation on the algebra.}

\endpage
\singlespace
\baselineskip=18pt

\chapter{\bf Introduction and Summary}

There has long been a desire reformulate string field theory in a simple
and concise form. It has already been seen in [\cbi], [\qbi], [\moduli],
[\nonconf] and [\basisone], how one might introduce the moduli spaces of
decorated Riemann surfaces $\B^{\bar n}_{g,n}$ of non-negative dimension
($g$ indicating the genus, $n$ the number of ordinary punctures and ${\bar n}$
the number of special punctures), which implement the
first and higher order background deformations of the closed bosonic string.
There is the hope however, not only of completing this set but also of
somehow including the remaining $\B$-spaces which can take {\it all}
non-negative integral values of $(g, n, \bar n)$, and in some way
incorporating these into the action. The main obstruction to this
had been the problem of interpreting those objects of `negative dimension'
which cannot be thought of as moduli spaces of surfaces in the usual
sense.

A step in this direction was achieved in the work [\nonconf]
of Zwiebach when the moduli space $\B^1_{0,1}$, na\"ively of
dimension\foot{Recall that the dimension of $\B^{\bar n}_{g,n}$ is
$6g-6+2n+3\bar n$.} $-1$, was introduced, being a sphere with one ordinary and
one special puncture. Moreover, the antibracket sewing of this
state was identified with the action of the operator $\I$ responsible for
changing an ordinary puncture into a special puncture, and the associated
function $f(\B^1_{0,1}) = - B_F^{(2)} = - \bra{\omega_{12}} F\rangle_1
\ket{\Psi}_2$, which acquires a ghost insertion, was absorbed into the string
action. This remarkable result demonstrated that the operator $\I$, usually
thought of as an outer derivation of the B-V algebra of string vertices,
could itself be represented by the action of one of the string vertices, thus
transforming it into an inner derivation of the algebra.
As a result he was able to simplify the recursion relations for the
$\B$-spaces, which took the following form,
$$\p \B - \K \B + \half \{ \B , \B \} - \V'_{0,3} = 0\,.\eqn\sabbytwo$$
In the above, $\B$ is a sum of moduli spaces of Riemann surfaces $\B =
\sum_{n,\bar n} \B^{\bar n}_{0,n}$ with $n$ ordinary punctures and $\bar n$
special punctures. In this summation $n$ and $\bar n$ can take all
non-negative integral values with the exception of the spaces
$\B^{\bar n}_{0,0}$ (which would contribute constant terms to the action),
$\B^0_{0,1}$ and $\B^0_{0,2}$. Given the success had with the operator $\I$,
one is led to ask whether the operators $\p$ and $\K$ may similarly be
expressible as inner derivations of the B-V algebra.

In this work we take the first step towards such a simplification. In
particular, we succeed in deriving
identifications for the operators $\p$, $\K$ and $\I$, the auxiliary vertices
$\V'_{0,3}$ and $\T^2_{0,1}$, and the Hamiltonians $Q$ and $B_F^{(2)}$ in
terms of the negative-dimensional spaces $\B^1_{0,1}$ and $\B^0_{0,2}$
(which we introduce). This implies the non-intuitive result that the
operators $\p$ and $\K$, previously thought to be outer derivations
having no obvious description in terms of string vertices, could possibly be
expressed as inner derivations. The recursion relations then take the form of
a quantum master action for the $\B$-spaces. The problem of simplifying the
action turns out to be much harder, and will be addressed in [\action].

\noindent
Our paper is organised as follows.

In \S 2.1 we introduce the moduli space $\B^0_{0,2}$ which has dimension $-2$
and is identified with the kinetic term $Q$ of the action, which is also the
Hamiltonian associated to the BRST operator. In \S 2.2 we list the operator
identities
satisfied by $\p$, $\K$ and $\I$. From these, we succeed in deriving the
unique set of requirements on the spaces $\B^0_{0,2}$ and $\B^0_{0,1}$
implying consistent operator identifications not only for $\p$, $\K$ and $\I$,
but also the vertices $\V'_{0,3}$ and $\T^2_{0,1}$. This demonstrates that
these operators may be expressed as inner derivations as we had hoped. We
also introduce related operators $\wt\p$, $\wt\K$ and $\H$.

In \S 2.3 we state the resulting form of the quantum action around arbitrary
string backgrounds and the corresponding recursion relations. These take a
completely geometrical form in that they are described by a quantum master
action for the string vertices. By expanding them we verify that they
agree precisely with the earlier form.

In \S 2.4 we attempt to express the B-V delta operator as an inner derivation
as was done for the other operators. This attempt fails as it has the
implication of a vanishing antibracket.

We end in \S 3 with the conclusion. Unless explicitly
stated otherwise, we will use units in which $\hbar = \kappa = 1$ throughout.

\chapter{\bf SFT Operators as Inner Derivations of the B-V Algebra}

In this section, we shall introduce a new moduli space $\B^0_{0,2}$
associated with the BRST Hamiltonian $Q$, and use it to attempt to identify
the action of the operators $\p$, $\K$ and $\I$ with the antibracket sewing of
elements of the B-V algebra of string vertices. The fact that we succeed shows
that these operators are expressible as inner derivations if the spaces
$\B^0_{0,2}$ and $\B^1_{0,1}$ satisfy certain unusual properties.

\section{The Brst Hamiltonian $Q$ and the Moduli Space $\B^0_{0,2}$}

The BRST Hamiltonian $Q = \half \bra{\omega_{12}} Q^{(2)} \ket{\Psi}_1
\ket{\Psi}_2$, which is also the kinetic term in the string action, may be
represented by the standard twice-punctured sphere with both a $c_0^-$
ghost and a BRST operator insertion as well as the two string field
insertions. In the spirit of [\nonconf], just as $B_F^{(2)} =
\bra{\omega_{12}} F \rangle_1 \ket{\Psi}_2$ was identified with
$- f(\B^1_{0,1})$, it is natural to introduce a moduli space $\B^0_{0,2}$
of dimension $-2$ which
will be the standard sphere with two ordinary punctures which we have just
mentioned, and then to identify $Q$ with $f(\B^0_{0,2})$. If we include
$\B^0_{0,2}$ in the sum\foot{Note for the quantum case that $\B$ also includes
the higher genus spaces of positive dimension introduced
in [\basisone].} $\B = \sum_{g,n,\bar n} \B^{\bar n}_{g,n}$, this
will allow us to absorb the BRST Hamiltonian into the term $f(\B)$ so that the
expression for the quantum action simply becomes,
$$S = S_{1,0} + f(\B)\,,\eqn\sabbyfive$$
where $S_{1,0}$ is the one-loop vacuum term, not present in the classical
theory.

We state an identity which we will make use of in the next section. Let
$\A$ be any moduli space. Then from Eqn.(3.25) of [\nonconf],
$$\{ Q , f(\A) \} = - f( \p \A - (-)^{\bar\A} \K \A)\,,\eqn\sabbythree$$
where have introduced for convenience the notation $\bar\A \equiv \A + \bar
n_\A$ for the grading. Given that $Q = f(\B^0_{0,2})$ this may be written,
$$f( \{ \B^0_{0,2} , \A \} ) = f( \p \A - (-)^{\bar\A} \K \A)
\,.\eqn\sabbysix$$
This will be useful in finding moduli space identifications for the operators
$\p$ and $\K$, to which we now turn.

\section{$\p$, $\K$ and $\I$ - Explicit Operator Identifications}

We will now suppose that the operators $\p$, $\K$ and $\I$ can be written as
inner derivations on the B-V algebra and shall attempt to derive the general
expressions for these operators in terms of elements of the B-V algebra of
string vertices, using the standard operator identities satisfied by them
as constraints. We recall the following list of operator identities,
$$\eqalign{\p \{ \A , \B \} &= \{ \p \A , \B \} + (-)^{\bar\A+1} \{ \A
, \p \B \}\,,\cr
\K \{ \A , \B \} &= (-)^{\bar\B+1} \{ \K \A , \B \} + \{ \A ,
\K \B \}\,,\cr
\I \{ \A , \B \} &= (-)^{\bar\B+1} \{ \I \A , \B \} + \{ \A ,
\I \B \}\,.}\eqn\sabfiftytwofa$$
$$\p^2 = \K^2 = \I^2 = 0\,.\eqn\sabfiftytwofb$$
$$\eqalign{(\K \I + \I \K) \A &= \{ \A , \T^2_{0,1} \}\,,\cr
[ \p , \I ] \A &= 0\,,\cr
[ \p , \K ] \A &= (-)^{\bar\A} \{ \V'_{0,3} , \A \}
\,.}\eqn\sabfiftytwofc$$
We also have the following properties, the first being Eqn.\sabbysix, and the
second being Eqn.(5.17) of [\nonconf],
$$\eqalign{f(\p \A &- (-)^{\bar\A} \K \A) = - \{ Q , f(\A) \} =
f(\{ \B^0_{0,2} , \A \})\,,\cr
&f(\I \A) = \{ f(\A) , B_F^{(2)} \} = f (\{ \A , \B^1_{0,1} \})\,.}
\eqn\sabfiftytwofca$$
The three conditions which needs be satisfied to write string theory around
non-conformal backgrounds Eqns.(3.9)-(3.11) of [\nonconf], are as follows,
$$\{ Q , B_F^{(2)} \} = - f(\T^2_{0,1})\,,\eqn\sabfiftytwogo$$
$$\half \{ Q , Q \} = f(\V'_{0,3})\,,\eqn\sabfiftytwogp$$
$$\{ Q , f(\Sigma) \} = f(\K \Sigma)\,.\eqn\sabfiftytwogq$$
The third of these is already satisfied as it was used to derive the first of
Eqns.\sabfiftytwofca. The remaining identities not included in the list above
are as follows,
$$\p \T^2_{0,1} = \I \V'_{0,3}\,,\eqn\sabfiftytwogqc$$
$$\T^2_{0,1} = \K \B^1_{0,1}\,,\eqn\sabfiftytwogqb$$
$$\V'_{0,3} = \K \B^0_{0,2}\,.\eqn\sabfiftytwogqd$$
The first is Eqn.(4.18) of [\moduli], the second is Eqn.(5.18) of [\nonconf],
the third was explained in \S 3.4 of [\nonconf], and the fourth was
mentioned as a postulate in a footnote in \S 5.3 of [\nonconf].
Our aim will be to find operator identifications and conditions on
$\B^0_{0,2}$ and $\B^1_{0,1}$ which satisfy all of these equations.

Let us first attempt to satisfy Eqns.\sabfiftytwofa. We shall consider
sums of operators of the form $\{ \X , \A \} = - (-)^{(\bar\A+1)(\bar\X+1)}
\{ \A , \X \}$, and $\{ \A , \Y \} = - (-)^{(\bar\A+1)(\bar\Y+1)} \{ \Y , \A
\}$, where $\X$ and $\Y$ are elements of the B-V algebra.
So we shall try the general forms,
$$\eqalign{\p \A &= \{ L , \A \} + \{ \A , \R \}\,,\cr
\K \A &= - \{ \L , \A \} - \{ \A , R \}\,,\cr
\I \A &= \{ r , \A \} + \{ \A , l \}\,,}\eqn\sabfiftytwofcaa$$
where $L$, $R$, $\L$, $\R$, $l$ and $r$ are elements of the B-V algebra to be
found. The reason for the unusual notational and sign choices will be
evident soon.

Let us consider the action of the general operator $\O \A = \{ \X , \A \} +
\{ \A , \Y \}$ on the antibracket, and use it to deduce conditions on our
unknown elements. We find,
$$\eqalign{\O \{ \A , \B \} &= \{ \X , \{ \A , \B \} \} + \{ \{ \A , \B \} ,
\Y \}\cr
&= - (-)^{(\bar\B+1)(\bar\X+1)} \bigl( (-)^{(\bar\A+1)(\bar\B+1)} \{ \B ,
\{ \X , \A \} \} + (-)^{(\bar\A+1)(\bar\X+1)} \{ \A , \{ \B , \X \} \} \bigr)
\cr
&\qquad - (-)^{(\bar\A+1)(\bar\Y+1)} \bigl( (-)^{(\bar\B+1)(\bar\Y+1)} \{ \{
\Y , \A \} , \B \} + (-)^{(\bar\A+1)(\bar\B+1)} \{ \{ \B , \Y \} , \A \}
\bigr)\cr
&= \{ (-)^{(\bar\B+1)(\bar\A\bar\X)} \{ \X , \A \} + (-)^{(\bar\B+1)
(\bar\Y+1)} \{ \{ \A , \Y \} , \B \}\cr
&\qquad + \{ \A , (-)^{(\bar\A+1) (\bar\X+1)} \{ \X , \B \} + (-)^{(\bar\A+1)
(\bar\B\bar\Y)} \{ \B , \Y \} \}\,.}\eqn\sabfourteen$$
We apply this to the operators $\p$, $\K$ and $\I$ whose postulated forms are
written in Eqn.\sabfiftytwofcaa, and compare with Eqn.\sabfiftytwofa. We come
to the following conclusions. For $\p$, we find that $L$ is graded-even, and
$\R$ vanishes. For $\K$, we find that $R$ is graded-even and $\L$ vanishes.
Finally, for $\I$ we find that $l$ is graded-even, and $r$ vanishes.

\noindent
To summarise, we have $\L = \R = r = 0$ and,
$$\eqalign{\p \A &= \{ L , \A \}\,,\cr
\K \A &= - \{ \A , R \}\,,\cr
\I \A &= \{ \A , l \}\,,}\eqn\sabfiftytwofda$$
where $L$, $R$ and $l$ are each graded-even elements of the B-V algebra.

\noindent
Consider now the nilpotency conditions. These imply the following
constraints on $L$, $R$ and $l$,
$$\{ L , L \} = \{ R , R \} = \{ l , l \} = 0\,.\eqn\sabfiftytwofdb$$
Let us now tackle the third set of conditions, Eqns.\sabfiftytwofc. For $\K \I
+ \I \K$ we find,
$$\eqalign{(\K \I + \I \K) \A &= - \{ \{ \A , l \} , R \} - \{ \{ \A , R \} ,
l \}\cr
&= \{ \{ l , R \} , \A \} + (-)^{\bar\A} \{ \{ R , \A \} , l \}
- \{ \{ \A , R \} , l \}\cr
&= - \{ \A , \{ l , R \} \}\,,}\eqn\sabfiftytwofq$$
which is of the form we seek. We may make the operator identification,
$$\T^2_{0,1} = - \{ l , R \}\,,\eqn\sabfiftytwofr$$
which must be non-vanishing. Turning now to the vanishing of $[ \p , \I ]$, we
find the condition,
$$\{ L , l \} = 0\,.\eqn\sabfiftytwofs$$
A similar calculation for $[ \p , \K ]$ implies the operator identification,
$$\V'_{0,3} = - \{ L , R \}\,,\eqn\sabfiftytwofu$$
which must also be non-vanishing. The fourth set of conditions,
Eqn.\sabfiftytwofca\ helps us to identify the vertices which $L$, $R$ and $l$
are associated with.  In particular we obtain the conditions,
$$L + R = \B^0_{0,2}\,,\eqn\sabfiftyofw$$
$$l = \B^1_{0,1}\,.\eqn\sabfiftytwofx$$
The latter condition fixes the form of $\I$ to be that stated in [\nonconf].
Consider now the conditions for string theory around non-conformal
backgrounds. Eqn.\sabfiftytwogo\ requires that,
$$\{ \B^0_{0,2} , \B^1_{0,1} \} = \{ l , R \}\,,\eqn\sabfiftytwoga$$
Similarly, Eqn.\sabfiftytwogp\ requires that,
$$\half \{ \B^0_{0,2} , \B^0_{0,2} \} = \{ L , R \}\,.\eqn\sabfiftytwogb$$
Finally, we consider the remaining identities,
Eqns.\sabfiftytwogqc-\sabfiftytwogqd. These imply the following set of
constraints,
$$\{ L , \{ l , R \} \} = \{ \{ L , R \} , l \}\,,\eqn\sabfiftytwogc$$
$$\{ l , R \} = \{ \B^1_{0,1} , R \}\,,\eqn\sabfiftytwogd$$
$$\{ L , R \} = \{ \B^0_{0,2} , R \}\,,\eqn\sabfiftytwogf$$
It is a remarkable fact that this entire gamut of constraints is soluble,
and that the unique solution is given by the following two requirements,
$$\eqalign{\B^0_{0,2} &= L + R\,,\cr
\B^1_{0,1} &= l\,,}\eqn\sabfiftytwogf$$
supplemented with the following conditions,
$$\{ L , L \} = \{ R , R \} = \{ l , l \} = \{ L , l \} = \L = \R = r = 0
\,.\eqn\sabfiftytwogg$$
We shall suggest an explanation for these conditions in a forthcoming
paper [\manifest]. These imply the set of operator identifications
Eqns.\sabfiftytwofda,
$$\eqalign{\p \A &= \{ L , \A \}\,,\cr
\K \A &= - \{ \A , R \}\,,\cr
\I \A &= \{ \A , l \}\,,}\eqn\sabfiftytwogj$$
and the special vertex identifications,
$$\eqalign{\V'_{0,3} &= - \{ L , R \}\,,\cr
\T^2_{0,1} &= - \{ l , R \}\,.}\eqn\sabfiftytwogl$$
For later use, we shall complete this set by introducing the following
additional trio of operators,
$$\eqalign{\wt\p \A &= \{ R , \A \}\,,\cr
\wt\K \A &= - \{ \A , L \}\,,\cr
\H \A &= - \{ l , \A \}\,.}\eqn\sabfiftytwogo$$
The operators $\wt\p$ and $\wt\K$ will be discussed in detail in [\manifest].
The operator $\H$ is really just the operator $\I$ with an additional
$\A$-dependent sign factor.
The success we have had in reproducing the usual operator identities
verifies that the operators $\p$ and $\K$ can, as was
shown to be the case in [\nonconf] for $\I$, be described as inner derivations
of the B-V algebra.

\section{Recursion Relations and the String Action}

Given all the preparatory work of the last section, we will now simply state
the form of the recursion relations and the string action around general
backgrounds and then check that they are correct. The recursion relations
are given by a `geometrical' quantum B-V master equation for the $\B$-space
complex,
$$\half \{ \B , \B \} + \Delta \B = 0\,,\eqn\sabfiftyfive$$
and the string action is given by Eqn.\sabbyfive,
$$S = S_{1,0} + f(\B)\,.\eqn\sabfiftysix$$
Here we have $\B = \sum_{g,n,\bar n} \B^{\bar n}_{g,n}$, where the sum is
over $\B$-spaces with all values of $(g,n,\bar n)$ except $\B^1_{1,0}$,
$\B^0_{1,0}$, $\B^{\bar n}_{0,0}$ and those
with $g=0$ and $n+\bar n\leq1$. That this action
satisfies the master equation is clear in view of the form of the
recursion relations and the field-independence of $S_{1,0}$.
For a background which is conformal, this reduces to the correct form
$S = Q + S_{1,0} + f(\V)$ where the kinetic term is $Q = f(\B^0_{0,2})$.

In order to gain some insight into how the recursion relations of
Eqn.\sabfiftyfive\ work, we will re-express them in such a way as to make them
look a little more familiar. Let us rewrite the negative-dimensional
$\B$-spaces in their operator form and let us also introduce the notation
$\B \equiv \B^0_{0,2} + \B^1_{0,1} + \wh\B = L + R + l + \wh\B$, where the
object $\wh\B$ is the restriction of $\B$ to the non-negative dimensional
$\B$-spaces. Then the recursion relations may be written as follows,
$$\eqalign{0 &= \half \{ \B , \B \} + \Delta \B\cr
&= \{ L , R \} + \{ l , R \} + \{ L + R , \wh\B \} + \{ l , \wh\B \} + \half
\{ \wh\B , \wh\B \} + \Delta \wh\B\cr
&= - \V'_{0,3} - \T^2_{0,1} + (\p - \K) \wh\B + \I \wh\B + \half
\{ \wh\B , \wh\B \} + \Delta \wh\B\,,}\eqn\sabfiftyseven$$
or perhaps more provocatively as,
$$\p \wh\B = \V'_{0,3} + \T^2_{0,1} + \K \wh\B
- \I \wh\B - \half \{ \wh\B , \wh\B \} - \Delta \wh\B\,,\eqn\sabfiftyeight$$
where in the first line we have expanded $\B$ in terms of its
components and used the fact that vertices are all
graded-even to extract the operators $\p$, $\K$ and $\I$.
Eqn.\sabfiftyeight\ should be compared with Eqn.(5.10) of [\nonconf]
\foot{Recall that $\M \equiv \K - \I$ and that the $\Delta$-term would be
expected in the na\"ive
quantum generalisation of Eqn.(5.10).}. The equivalence between our form of
the recursion relations and (the quantum generalisation of) those claimed by
Zwiebach is now transparent.

We might mention here that we have tacitly used the fact that
$\Delta L$ and $\Delta R$ both vanish in deriving
Eqn.\sabfiftyseven, which is in accord with $\Delta Q = 0$ [\csft]. A
geometrical explanation for this will also be suggested in [\manifest].

Before we end, we shall discuss briefly the possibility of expressing the
B-V delta operator as an inner derivation.

\section{The B-V Delta Operator - An Inner Derivation?}

The B-V delta operator satisfies several identities similar to those of
$\p$, $\K$ and $\I$. We have shown in \S 2 that these three operators may
consistently be expressed as inner derivations in such a way that the usual
identities are satisfied. We now ask ourselves the question of whether the
same might be possibly with the $\Delta$ operator. This seems an unlikely
proposal, but it turns out to be surprisingly close to working,
the failure being a somewhat indirect implication of the B-V antibracket
being a derivation of the graded-commutative and associative dot product on
moduli spaces.

\noindent
Let us list the usual identities known to be satisfied by the $\Delta$
operator,
$$\Delta \{ \A , \B \} = \{ \Delta \A , \B \} + (-)^{\bar\A+1} \{ \A , \Delta
\B \}\,,\eqn\sabsixtythree$$
$$(\p \Delta + \Delta \p) = 0\,,\eqn\sabsixtyfour$$
$$[ \Delta , \K ] = 0\,,\eqn\sabsixtyfoura$$
$$\Delta^2 \A = 0\,.\eqn\sabsixtyfive$$
Assuming the identifications derived earlier for $\p$, $\K$ and $\I$, the
second identity actually follows from the first which we demonstrate as
follows,
$$\Delta \p \A = \Delta \{ L , \A \} = \{ \Delta L , \A \} - \{ L , \Delta \A
\} = - \{ L , \Delta \A \} = - \p \Delta \A\,,\eqn\sabsixtysix$$
while a similar calculation verifies the third identity. We might also use
this construction to `predict' the identity $[ \Delta , \I ] = 0$.

Now Eqn.\sabsixtythree\ is very reminiscent of an analogous identity,
Eqn.\sabfiftytwofa\ which is satisfied by $\p$, and this strongly suggests an
identification,
$$\Delta \A = \{ \X , \A \}\,.\eqn\sabsixtyseven$$
where $\X$ is an element of the B-V algebra to be found. The fourth condition
of nilpotency gives the following condition on $\X$,
$$\{ \X , \{ \X , \A \} \}\,.\eqn\sabfifteen$$
This vanishes identically only if $\X$ is graded-even and satisfies,
$$\{ \X , \X \} = 0\,.\eqn\sabsixteen$$
This all seems very encouraging as the logic so far has been identical to
the cases of $\p$, $\K$ and $\I$. However, the B-V delta operator satisfies
one more identity, Eqn.(3.14) of [\bistruct], which relates it to (and can be
used to define) the B-V antibracket,
$$\{ \A , \B \} = (-)^{\bar\A} \Delta (\A \cdot \B) + (-)^{\bar\A+1} (\Delta
\A) \cdot \B - \A \cdot (\Delta \B)\,,\eqn\sabseventeen$$
where the moduli spaces form an algebra under the graded-commutative and
associative product denoted by the `$\, \cdot \,$'. Applying
Eqn.\sabsixtyseven\ to Eqn.\sabseventeen, we find the following condition,
$$\{ \A , \B \} = (-)^{\bar\A} \{ \X , \A \cdot \B \} + (-)^{\bar\A+1} \{ \X ,
\A \} \cdot \B - \A \cdot \{ \X , \B \}\,.\eqn\sabeighteen$$
But the antibracket satisfies the following property with respect to the
dot product,
$$\{ \X , \A \cdot \B \} = \{ \X , \A \} \cdot \B + (-)^{(\bar\X+1)\bar\A} \A
\cdot \{ \X , \B \}\,.\eqn\sabnineteen$$
Our space $\X$ is graded-even, so applying this to Eqn.\sabeighteen\ we
find that this would imply that the antibracket identically vanishes! This
shows that the B-V delta operator cannot be described as an inner derivation
of the algebra, despite initial promise.

\chapter{\bf Conclusion}

We have shown that it is possible to consistently identify the usual string
field theory operators with the elementary moduli spaces $\B^0_{0,2}$
and $\B^1_{0,1}$ of `negative
dimension', so that $\p$, $\K$ and $\I$, may be considered as inner
derivations of the B-V algebra of string vertices.

The recursion relations took the form of a `geometrical' quantum master
equation, a fact which suggests that this a phenomenon restricted not only to
string theory, but may be of much more general applicability. The action
is still not quite in the form we would like, namely $S=f(\B)$, and this is a
reflection of the fact that there are still several string vertices excluded
from the sum $\B$. Including the remaining string vertices proves to be
difficult yet possible, and this will be the subject of a forthcoming paper
[\action].

An attempt to express the operator $\Delta$ as an inner derivation on the B-V
algebra failed, though the matter shall be discussed again under different
circustances in [\manifest].

\ack
I would like to thank my supervisor Barton Zwiebach, and am grateful to Chris
Isham for allowing me to attend Imperial College, London, where much of this
work was completed. I would also like to extend my sincere apologies to Robert
Dickinson.

\refout

\bye